\documentclass[12pt]{article}
\usepackage{a4wide,graphicx,amsmath,cite,longtable,multicol,array}

\makeatletter
\@addtoreset{equation}{section}

\makeatother

\newcommand{\beq}{\begin{equation}}
\newcommand{\eeq}{\end{equation}}
\newcommand{\beqa}{\begin{eqnarray}}
\newcommand{\eeqa}{\end{eqnarray}}

\begin{document}

\begin{titlepage}

\hfill\begin{tabular}{l}
UG-FT-254/09 \\ 
CAFPE-124/09 \\ 
RM3-TH/09-1
\end{tabular}

\bigskip
\bigskip
\bigskip
\bigskip

\begin{center}

\begin{Large}
\textbf{\boldmath{Atmospheric lepton fluxes at ultrahigh energies}}
\end{Large}

\bigskip
\bigskip
\bigskip
\bigskip

{\large 
Jos\'e Ignacio Illana$^a$, Manuel Masip$^a$, Davide Meloni$^b$}~\footnote{%
        e-mails: jillana@ugr.es, masip@ugr.es, meloni@fis.uniroma3.it}

\bigskip
\bigskip

\begin{it}
$^a$Depto. de F{\'\i}sica Te\'orica y del Cosmos, Universidad de Granada,
18071 Granada, Spain

\smallskip

$^b$Dipto. di Fisica, Universit\`a di Roma Tre, 00146 Rome, Italy
\end{it}

\bigskip
\bigskip
\bigskip
\bigskip

\textbf{Abstract}

\end{center}

\noindent
In order to estimate the possibility to observe exotic 
physics in a neutrino telescope, it is essential 
to first understand the flux of atmospheric neutrinos, 
muons and dimuons.
We study the production of these leptons by high-energy 
cosmic rays. We identify three main sources of muons
of energy $E\ge 10^6$ GeV: the {\it weak} 
decay of charm and bottom mesons and the {\it electromagnetic}
decay of unflavored mesons. Contrary to the standard assumption, 
we find that $\eta$ mesons, {\it not} the prompt decay of charm 
hadrons, are the dominant source of atmospheric 
muons at these energies. 
We show that, as a consequence, the ratio between the
neutrino and muon fluxes is significantly reduced.
For dimuons, which may be a background 
for long-lived staus produced near a neutrino telescope, 
we find that pairs of $E\approx 10^7$ GeV 
forming an angle above $10^{-6}$ rad 
are produced through $D$ (80\%) or $B$ (10\%) meson
decay and through Drell-Yan proceses (10\%).
The frequency of all these processes 
has been evaluated using the jet code PYTHIA.

\end{titlepage}

%%%%%%%%%%%%%%%%%%%%%%%%%%%%%%%%%%%%%%%%%%%%%%%%%%%%%%%%%%%%%%%%%%%%%%%%

\section{Introduction}

We observe a flux of cosmic rays (protons free or 
bound in nuclei) that extends up to energies of  
$10^{11}$ GeV \cite{Amsler:2008zzb}. When a very energetic 
cosmic ray enters the atmosphere
it will collide with a nucleus of air at an 
approximate altitude of 20 km. The interaction 
breaks the primary proton, starting
an air shower that includes millions of particles,
collisions and decays. The development of the
shower along the atmosphere (number of particles
and energy deposited at different depths) is fairly well 
understood \cite{Greisen:1960wc,Lipari:1993hd}. 
In particular, computer simulations 
provided by codes like AIRES \cite{aires}
or CORSIKA \cite{Heck:1998vt} reproduce 
well the profile of the shower and the number of 
charged particles (mostly muons) that reach the ground.

Most of the processes that take place inside the air
shower are {\it soft} hadronic collisions and particle
decays. Typically, a hadron is initially 
produced at a given point in the atmosphere with a 
large boost towards
the ground. If this hadron is a proton or a neutron
it will move along a hadronic interaction length 
($\lambda_{\rm int} \approx 1$--6 km, depending on the altitude) 
and will collide with a nucleus, 
exchanging momentum with $q^2<1\;{\rm GeV}^2$. 
The process will result into a leading hadron plus 
several other hadrons sharing the total energy. 
On the other hand, if the initial hadron 
is a charged pion (or a kaon) the sequence of 
events depends critically on its energy $E$. For $E$ larger 
than $E_{\rm crit}^{\pi}=115$ GeV \cite{Costa:2000jw}
(or $E_{\rm crit}^{K}=855$ GeV) its interaction length is 
smaller than its decay length, the pion tends to collide 
and behaves much like the stable proton. \footnote{The actual value of $E_{\rm crit}^{\pi}$ depends
on the air density at the point where the pion is produced.}
For smaller 
energies, however, it decays and produces the 
{\it conventional} muon and neutrino fluxes.
Finally, a third possibility is that the initial hadron 
decays into other hadrons, photons or
charged particles {\it very fast} through strong or
electromagnetic interactions (this is the case for
resonances, neutral
pions, and other hadrons that do not decay {\it weakly}).

Some physical observables, however, may depend on
{\it rare} processes of higher $q^2$ 
that also occur inside the shower 
and cannot be overlooked. For example, consider 
the atmospheric muon flux above $10^6$ GeV. At these 
energies charged pions collide with the air before 
they can decay. The origin of these 
muons, as it has been widely described in the
literature \cite{Volkova:1987gh,Bugaev:1998bi,Gondolo:1995fq,
Zas:1992ci,Pasquali:1998ji,Gelmini:1999xq,Enberg:2008te,
Kochanov:2009rn}, is the
prompt decay of charmed hadrons produced in hadron-nucleon
collisions. Charm production involves $q^2>(2m_c)^2$ and
is less frequent than pion or kaon production. One may
argue that the number of these very energetic muons 
is negligible compared to the total number of muons 
reaching the ground, and also that the energy that they 
take from the shower is a non-significant fraction
of the total energy deposited in the atmosphere. Such 
arguments 
could justify the absence of charm and bottom hadrons 
in the air-shower simulations performed by AIRES or CORSIKA. 
Even if negligible there, however,
these muons may be observable in a neutrino telescope
like IceCube \cite{icecube,Berghaus:2009jb}, 
specially from near-horizontal directions.
Notice that given a zenith angle $\theta$ 
fixing the ice column density faced by an atmospheric
particle in its way to the telescope, only muons 
above some energy threshold can reach the 
detector. An IceCube measurement of the muon flux 
at $E\ge 10^6$ GeV would provide information
about hadronic collisions which is 
complementary to collider data, as this {\it forward} 
physics at such high energies is not available there.

In this paper we evaluate the production of high-energy 
atmospheric leptons: muons, neutrinos and dimuons. 
We are interested in the energy region around $10^7$ GeV, 
where a $D$ meson tends to interact before decaying just 
like a pion does already at lower energies.
Our interest is motivated
by several factors. The atmospheric muon and neutrino 
fluxes are correlated, so a measurement of the former
determines the latter. Ultrahigh energy atmospheric 
neutrinos are of interest 
because {\it (i)} they are a
background to a possible cosmic neutrino flux to be 
observed in IceCube, and  {\it (ii)} it has been suggested
\cite{Ando:2007ds} that they could be a source of new 
physics ({\it e.g.,}
$\nu N\rightarrow \tilde\tau\tilde\tau X$) 
observable at a neutrino telescope. 
Moreover, atmospheric muon pairs from quasi-horizontal 
directions are themselves a 
background to exotic new physics at a neutrino telescope. 
For example, they could be confused with a pair of 
long-lived charged massive particles (CHAMPs)
produced near the 
detector through a $\nu N$ interaction \cite{Ahlers:2007js}.

We think that previous estimates of the muon
and neutrino fluxes at ultrahigh energies have overlooked
some very relevant effects. Most notably, they do not
include muons produced in the electromagnetic decay of 
unflavored mesons. In addition, some of them 
do not properly include
the propagation in the atmosphere of charmed
hadrons of $E> 10^7$ GeV (see below).
Finally, we include in our study the Drell-Yan process 
$q\bar{q}\rightarrow \gamma/Z \rightarrow \mu\mu$, which
cannot be neglected when studying the background
to exotic physics at neutrino telescopes.

\section{Atmospheric muon and neutrino fluxes}

We will focus on leptons of energy above $10^6$ GeV. 
As explained before, at these energies
the contribution from charged-pion 
decay is negligible. Actually, we will take secondary 
pions and kaons at these energies as stable
particles that may produce muons in their
collision (not decay) 
with a nucleus of air. In \cite{Illana:2006xg}
we find an estimate of the total atmospheric flux of 
primary plus secondary 
nucleons ($N$) and long-lived mesons ($\pi^\pm$ and
$K$). 
We have used the jet code 
PYTHIA \cite{pythia} to simulate their
collisions and identify the main sources of leptons,
which are the following.

\vskip 0.2truecm
\noindent {\it (i)} The {\it standard} source of
muons of energy above $10^6$ GeV is 
the prompt decay of charmed hadrons. 
These hadrons decay into muons 
with a branching ratio of about $10\%$.

\vskip 0.2truecm
\noindent {\it (ii)} As the energy of a charmed meson
grows, its decay length $\lambda_{\rm dec}$
becomes larger than its interaction 
length $\lambda_{\rm int}$, 
reducing the probability of decay into 
leptons before colliding. 
The possibility that the charmed
hadron interacts several times in the air 
and then decays, however, can not be neglected. 
The main reason is that a $D$ meson 
does {\it not} propagate in the air like a pion. The 
charmed hadron is much more penetrating. We estimate that
in a hadronic interaction it loses just a fraction
\beq
x\equiv \frac{\Lambda}{m_c}\approx 0.3
\eeq
of the energy lost by a pion of equal energy. After
each interaction with air there is {\it always}
a leading charmed hadron carrying more than $70\%$ of
the initial energy. 
We find, for example, that the decay
of charmed hadrons that have interacted with
the atmosphere before increases the prompt lepton flux 
at $E\ge 10^7$ GeV by 30\%.
Previous analyses \cite{Gondolo:1995fq}
have assimilated the propagation
of a $D$ meson (or a $\Lambda_c$ baryon) to that of a 
pion (or a $\Lambda$), which reduces the 
energy of the leading hadron 
and makes this contribution negligible. Other studies
however, have 
included this effect properly (for example, through
a regeneration factor $Z_{c\overline c}=0.8$ in 
\cite{Martin:2003us}).

\vskip 0.2truecm
\noindent {\it (iii)} The two processes above would 
also work for bottom hadrons, which have similar lifetime
and semileptonic branching ratio.
Of course, $B$ mesons are less frequently 
produced than $D$ mesons by cosmic rays in the 
atmosphere, but their decay length is smaller.
Their contribution to the $10^7$ GeV muon flux
is a 10\% of the one from charm hadron decays.

\vskip 0.2truecm
\noindent {\it (iv)} The electromagnetic decay of 
unflavored hadrons, most notably $\eta$ mesons, has
been almost completely overlooked in the literature
(the only generic mention 
that we have found is given by Ryazhskaya, Volkova 
and Zatsepin in \cite{Ryazhskaya:2005rp}). Since these
hadrons do not contain any heavy quarks, they are more
abundant in air showers than charmed or bottom hadrons
($\eta$ mesons are produced through low $q^2$
interactions, just like pions or kaons).
In addition, since they decay through electromagnetic
processes their lifetime is much shorter than the one
of charged pions, kaons or $D$ mesons, whose decay is
mediated by weak interactions. The decay 
length of a $10^{10}$ GeV $\eta$ meson, for example, 
is just 280 m.
On the other hand, the preferred decay modes are into
photons and neutral pions, the branching fraction to 
$\mu^+\mu^-\gamma$ (around  $3.1\times 10^{-4}$)
being suppressed by an $\alpha_{QED}$ factor. 
In the lenguage of the Z-moment method, these 
production and decay rates would translate into 
an approximate $Z_{p\eta}\approx Z_{pK_L}$ and 
$Z_{\eta\mu}\approx 2\times 10^{-3} Z_{K_L\mu}$.
Notice also that, in contrast with the other (weak)
processes, these contributions break the correlation
between the atmospheric muon and neutrino fluxes.

\vskip 0.2truecm
\noindent {\it (v)} Finally, we will also include 
in our analysis the Drell-Yan process 
$q\overline q\rightarrow \gamma/Z \rightarrow \mu^+ \mu^-$.
This scattering will be relevant to determine  
the flux of muon pairs ({\it dimuons}) forming a
minimum angle such that they can be resolved at a neutrino
telescope.

%%%%%%%%%%%%%%%%%%%%%%%%%%%%%%%%%%%%%%%%%%%%%%%%%%%%%%%%%%%%%%%%%%%%%%%%%%%%%%%
\begin{figure}
\centerline{\includegraphics[width=8cm]{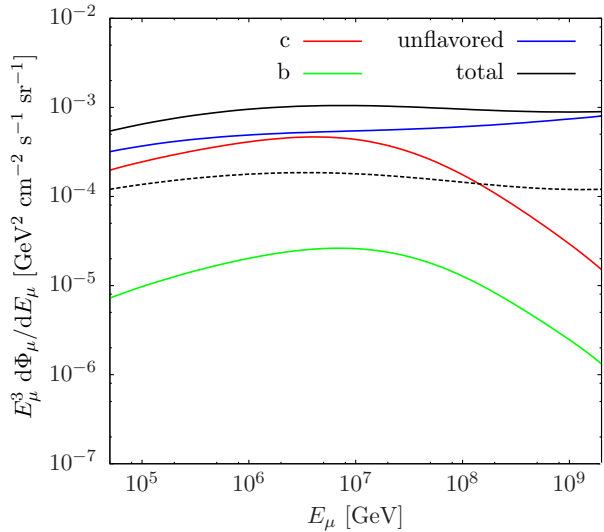}}
\caption{
Muon flux started by nucleon and meson collisions from the sources {\it (i--iv)} (solid) and just from meson collisions (dashes). Conventional muons from pion decay are not included. \label{fig1}}
\end{figure}
%%%%%%%%%%%%%%%%%%%%%%%%%%%%%%%%%%%%%%%%%%%%%%%%%%%%%%%%%%%%%%%%%%%%%%%%%%%%%%%

Our results are summarized in Figs.~1--2. We have 
simulated hadron ($h=p,n,\pi^+,K$ and antiparticles) 
collisions with atmospheric nucleons and identified the
muons and neutrinos produced through the five processes 
described above. 
We have estimated the probability for a
process $X$ by comparing its cross section $\sigma^{hN}_X$
(obtained from PYTHIA) 
with the total cross section with the air:
\beq
{\cal P}^h_X(E) \approx 
\frac{A \; \sigma^{hN}_X}{\sigma^{ha}_{T}}\;,
\label{eq3}
\eeq
where $A=14.6$ is the averaged atomic mass of a nucleus
of air.\footnote{
By using the factor $A$ instead of $A^{2/3}$
we ignore the screening between target nucleons and 
include the possibility that the incident hadron
collides with more than one nucleon inside the 
nucleus of air (its partons do not {\it disappear} 
after the first interaction).
As a consequence, (\ref{eq3}) expresses
the number of (short distance) interactions per collision 
with a nucleus, and could be larger than one.} 
PYTHIA has simulated
the production of five types of charmed hadrons
($D^0$, $D^+$, $D^+_s$, $\Lambda^+_c$, $\Omega^0_c$
and their antiparticles), seven of bottom
hadrons ($B^0$, $B^+$, $B_s^+$, $B_c^+$, $\Lambda_b^0$,
$\Xi_b^0$, $\Xi_b^+$) and three of unflavored mesons 
($\eta$, $\rho$ and $\phi$).
PYTHIA also takes care of gluon 
emission, the decay of hadronic
resonances, and the decay into leptons of the
parent heavy or unflavored hadron. 
These probabilities
have then been convoluted with the total fluxes in 
\cite{Illana:2006xg}.

We have estimated for all these hadrons 
an interaction length in the air of 5 km, allowing that
they decay between interactions (we have neglected the
variation of $\lambda_{dec}$ with the altitud; 
since most of these long-lived
hadrons are produced between the second the fourth
interaction lengths, the 5 km should be an acceptable
approximation).
We have assumed that 
in each interaction they lose 
30\% (charm) or 10\% (bottom) of their energy.

In Fig.~1 we plot the flux of muons
at different energies. We have separated the contributions
from the prompt decay of charmed hadrons 
(that may have interacted before
decaying) ({\it i--ii}), from $B$ decays ({\it iii}), 
and from $\eta$ decays ({\it iv}).
We have also separated the contribution from 
collisions of secondary mesons (charged pions and kaons) 
with air nucleons, which is about a 15\% of the total.

One important point concerns the procedure used 
to estimate the probability $p_{\rm dec}$ that, once produced, 
a hadron decays before interacting. 
PYTHIA assigns to the hadron a 
proper lifetime $\tau$
that is distributed around the average value $\tau_0$.
We boost $\tau$ and calculate the probability 
that in this time the charmed hadron does not interact with
the atmosphere before decaying:
\beq
p_{\rm dec}=\exp \left\{ -\frac{\tau E}{m \lambda_{\rm int}}\right\} \;,
\eeq
with $\lambda_{\rm int}\approx 5$ km.
Notice that always taking the average
value $\tau_0$ instead of $\tau$, 
the decay probability of a charmed meson 
of energy above $E_{\rm crit}= m \lambda_{\rm int}/\tau_0$
would be much smaller, since 
one would exclude the possibility of $\tau$ being 
significantly smaller
than $\tau_0$. We think this may be the reason why other
estimates predict a sharper drop in the muon flux at energies
above $10^8$ GeV. Other than that, the contribution from
charm production that we obtain with PYTHIA (based
on perturbative QCD) is similar 
to the one given in \cite{Gondolo:1995fq}.
Soft physics is treated by PYTHIA using reggeon and
pomeron exchange \cite{Donnachie:1992ny,Schuler:1993wr}
and the string fragmentation 
model \cite{Andersson:1983ia}.

%%%%%%%%%%%%%%%%%%%%%%%%%%%%%%%%%%%%%%%%%%%%%%%%%%%%%%%%%%%%%%%%%%%%%%%%%%%%%%%
\begin{figure}
\centerline{\includegraphics[width=8cm]{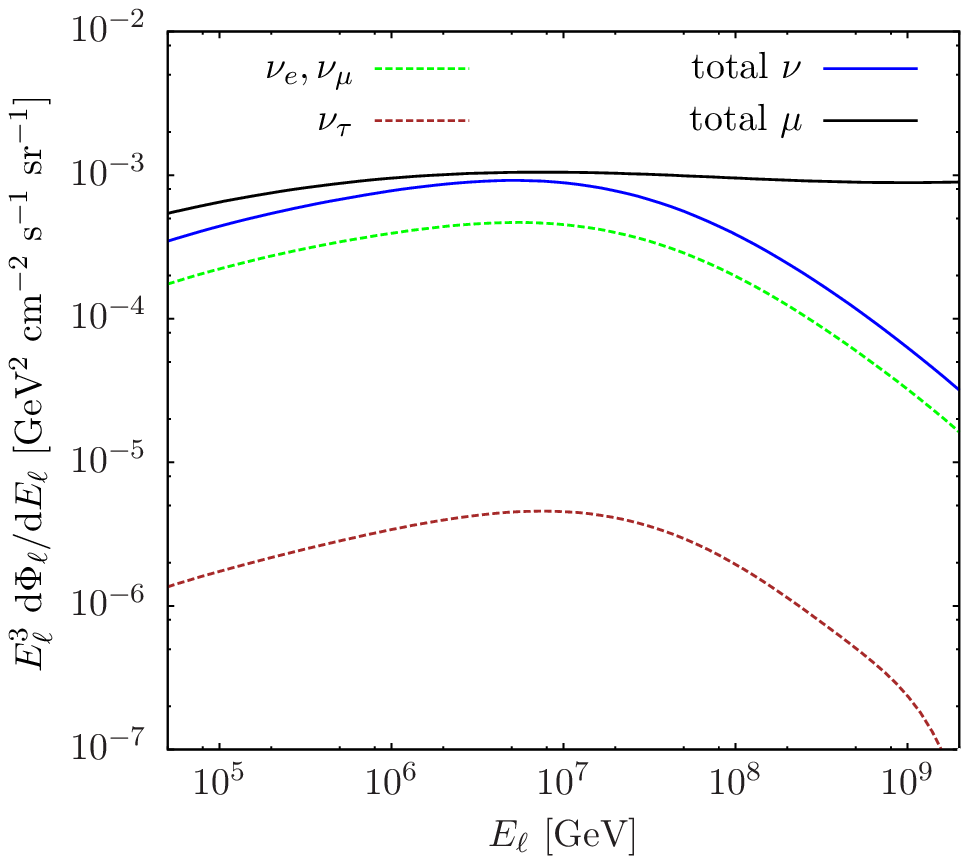}\quad
\includegraphics[width=8cm]{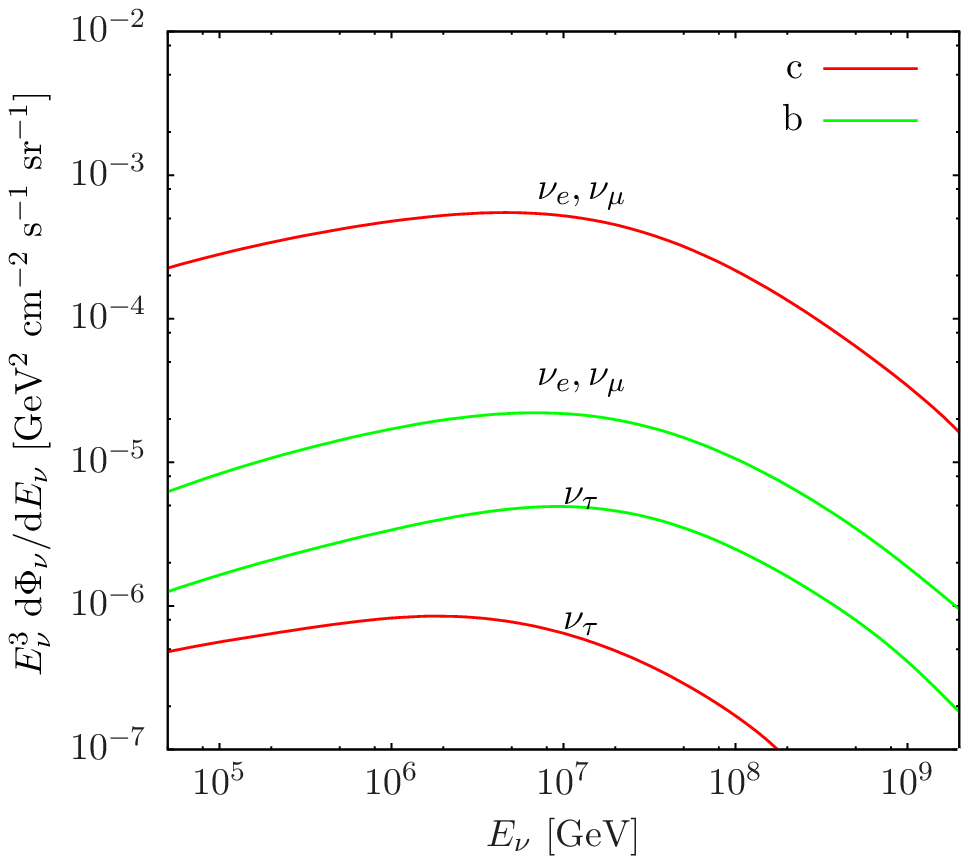}}
\caption{
Total muon and neutrino fluxes (left) and relative
contribution from charmed and bottom hadrons 
in the neutrino fluxes (right). 
\label{fig2}}
\end{figure}
%%%%%%%%%%%%%%%%%%%%%%%%%%%%%%%%%%%%%%%%%%%%%%%%%%%%%%%%%%%%%%%%%%%%%%%%%%%%%%%

In Fig.~2 (left) we show the total muon flux 
along with the fluxes 
of the three different neutrino species. The ratio of
muon neutrinos to muons changes from 1 to 0.2 at
$E=10^8$ GeV with the inclusion of the electromagnetic
decays of unflavored hadrons. In Fig.~2 (right) we
give the contribution of charmed and bottom decays
to the three neutrino species.
The contribution to the tau
neutrino flux from the prompt
decay of $D_s$ mesons that we obtain is
smaller than what one would expect
from a simple estimate (notice that
other charmed hadrons are too light to decay into
tau leptons). 
This result does not agree with previous analyses 
\cite{Martin:2003us,Pasquali:1998xf,Costa:2001fb,Lee:2004zm}, 
and seems to be caused by a too low decay rate 
of $D_s$ into $\tau \nu_\tau$ given by our 
PYTHIA simulation.\footnote{We have used the default 
PYTHIA 6.418 (fortan) distribution of the code.}

\section{Atmospheric flux of muon pairs}

Two muons separated
by a distance of 50 meters
crossing a neutrino telescope could 
look similar to a pair of CHAMPs.
If the CHAMPs come upwards \cite{Albuquerque:2006am}, 
their origin
could only be a neutrino interaction at some
distance from the detector: the upward 
atmospheric dimuon background vanishes  
whereas the typical distance (angle) between
two muons produced by a neutrino
near the detector is always smaller. 
However, if the CHAMPs reach the telescope
from a quasi-horizontal direction 
($\theta=70^\circ$--$95^\circ$),
they could have been produced in the atmosphere
(in a hadron-nucleon interaction) 
or in the Earth (a neutrino-nucleon collision), 
and the evaluation of the expected
number of atmospheric muon pairs becomes essential. 

We have estimated this flux at very
high energies $E_{\mu\mu}$. We require that the total energy  
is balanced, with each muon carrying at least $1\%$ of 
$E_{\mu\mu}$. This is necessary as {\it both} muons 
must be able to reach the telescope from a given distance.
Their origin are also the five processes
discussed in the previous section. Heavy quarks are 
always produced in pairs by cosmic rays, so it
may be that both $D$ (or $B$) mesons decay into muons
and define a dimuon.
The decay of an $\eta$ meson
gives dimuons (never single muons), but the transverse
momentum is very small and does not provide 
enough separation to resolve them. 
In addition, the Drell-Yan process becomes important
when a minimum angle between the two muons is imposed 
(notice that $\eta$ production and decay into muons, 
$q\bar{q}\rightarrow \eta\rightarrow \mu^+\mu^-$,
can be interpreted as a Drell-Yan at $q^2\approx m_\eta^2$).

%%%%%%%%%%%%%%%%%%%%%%%%%%%%%%%%%%%%%%%%%%%%%%%%%%%%%%%%%%%%%%%%%%%%%%%%%%%%%%%
\begin{figure}
\centerline{\includegraphics[width=8cm]{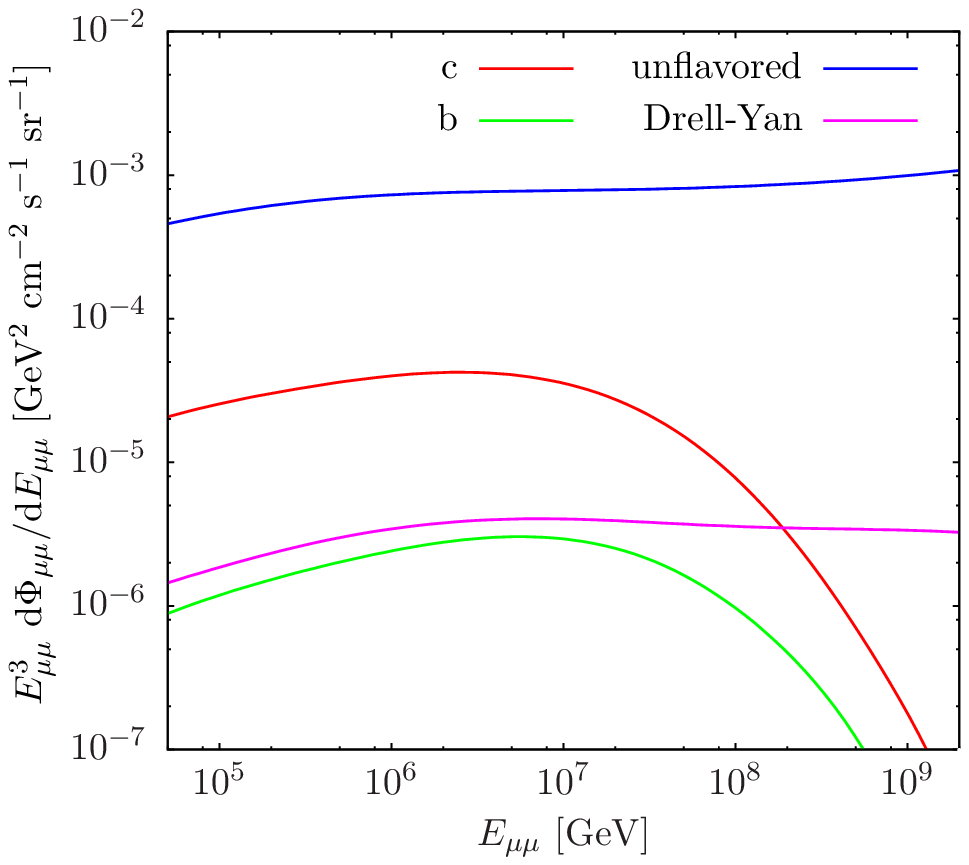}\quad
\includegraphics[width=8cm]{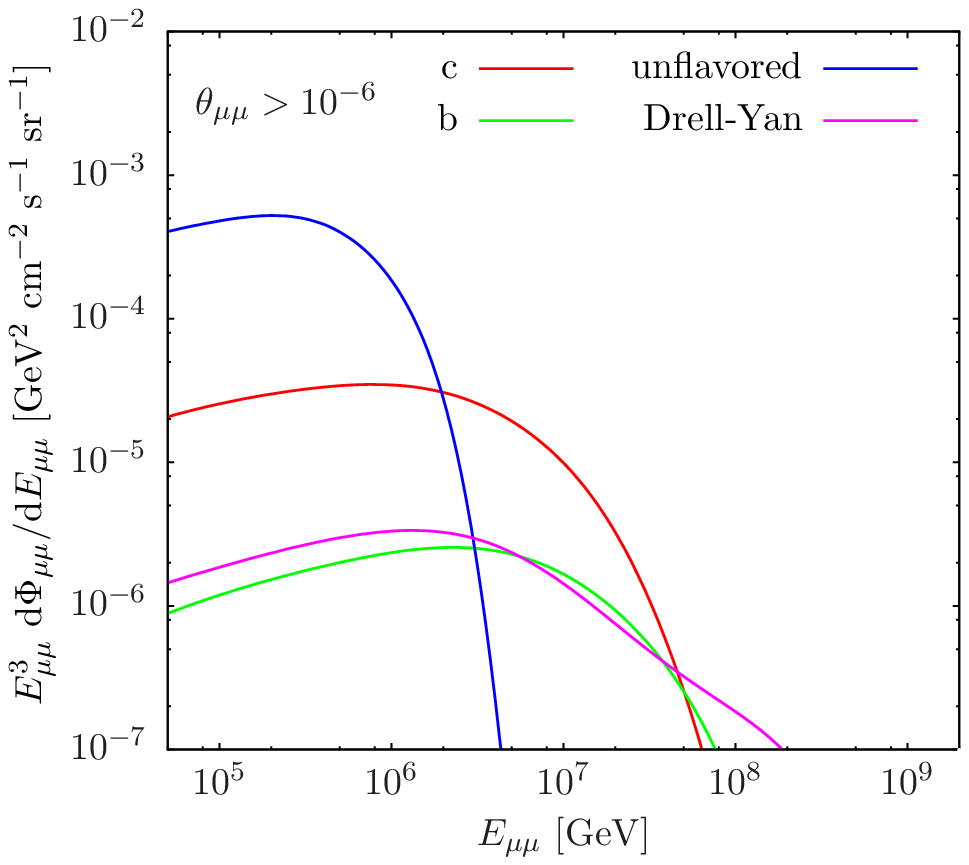}}
\caption{
Total dimuon flux (left) and dimuon flux 
requiring a minimum angle of 
$10^{-6}$ rad (right). We separate the
contribution from charm decay, bottom decay, unflavored
meson decay, and Drell-Yan processes.
\label{fig3}}
\end{figure}
%%%%%%%%%%%%%%%%%%%%%%%%%%%%%%%%%%%%%%%%%%%%%%%%%%%%%%%%%%%%%%%%%%%%%%%%%%%%%%%

In Fig.~3a--b we plot the total dimuon flux (left) and the flux 
requiring a minimum angle of  $10^{-6}$ rad (right). 
We separate the
contribution from charm decay, bottom decay, unflavored
meson decay, and Drell-Yan processes
(which include $\gamma$ and $Z$ exchange). It is apparent 
that the requirement of a minimum angle cuts all dimuons
from $\eta$ decays, while prompt leptons
dominate up to $10^7$ GeV and Drell-Yan at higher energies.
In Fig.~4 we give the total flux after cuts of
$0$, $10^{-8}$, $10^{-6}$ and $10^{-4}$ rad in the angle
between the two muons. 

%%%%%%%%%%%%%%%%%%%%%%%%%%%%%%%%%%%%%%%%%%%%%%%%%%%%%%%%%%%%%%%%%%%%%%%%%%%%%%%
\begin{figure}
\centerline{\includegraphics[width=8cm]{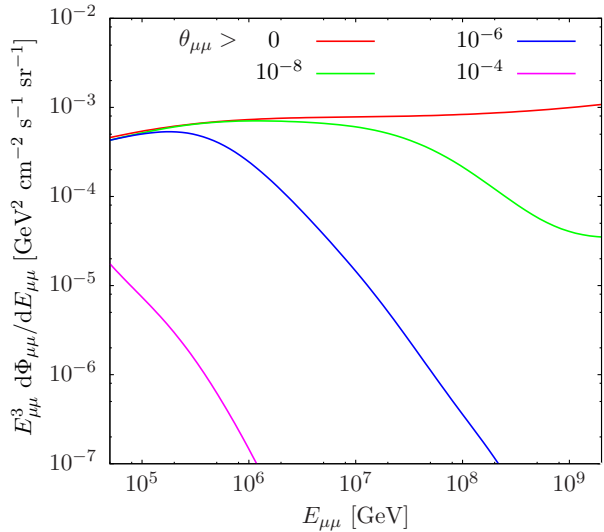}}
\caption{
Total dimuon flux for different cuts in their opening angle.
\label{fig4}}
\end{figure}
%%%%%%%%%%%%%%%%%%%%%%%%%%%%%%%%%%%%%%%%%%%%%%%%%%%%%%%%%%%%%%%%%%%%%%%%%%%%%%%

\section{Summary and discussion}

Cosmic rays produce 
extensive air showers with millions of
particles of different energy. The atmosphere acts as
a calorimeter and absorbs most of the initial energy.
However, a small fraction of this energy 
goes to very penetrating particles
that are able to
enter the ground and reach a neutrino telescope like
IceCube: muons and neutrinos.
If the detectors there {\it look} down, a signal could
only be due to neutrino interactions near (or inside) the 
telescope. The interacting neutrino could be atmospheric or 
cosmogenic, whereas the signal could correspond to 
standard (muons crossing the telescope upwards) or
exotic (stau pairs from the same direction). 
In any case, the determination of the
origin of the signal requires an accurate estimate
of the atmospheric neutrino flux at these energies
\cite{GonzalezGarcia:2006ay}.

In addition, if the telescope also covers
quasi-horizontal zenith angles ($\theta=70^\circ$--$95^\circ$)
there appear several new and interesting possibilities 
that could explain the origin of
a signal. The large ice column density along these
directions filters most of the components in an
air shower, just like from upward directions. 
Actually, everything but neutrinos and very
energetic muons will be absorbed by the ground. 
The experimental determination of the muon 
flux from these inclinations would be a direct
test for the different computer codes used to simulate
air showers, and also an indirect measurement of the
atmospheric neutrino flux at these energies.
More importantly, these horizontal directions are 
the most promising in the search for exotic
physics, since this physics may have been produced 
not only by cosmogenic or atmospheric neutrinos
in the ice, but also in muon-nucleon 
or hadron-nucleon collisions in the atmosphere.
For example, a couple of massive, long-lived 
(very penetrating) 
staus produced in the air via gluino 
decay would easily reach the center of a 
neutrino telescope from horizontal directions,
but not from below, as they would be unable to
cross the Earth. The identification
of this type of exotic physics requires an  
accurate estimate of the muon and dimuon 
atmospheric background.

We have determined the flux of muons and 
neutrinos at energies 
above $10^6$ GeV. We have found that $\eta$ decay,
not the prompt decay of $D$ mesons, is the main source 
of muons at these energies. Our results
confirm the estimate given in \cite{Ryazhskaya:2005rp}.
$D$ and $B$ mesons decay
via $W$ exchange and become long-lived at very high energies 
just like charged pions do at lower energies.
In contrast, $\eta$ mesons 
decay fast through electromagnetic interactions
(just like neutral pions). The $\eta$ 
branching ratio to muons is suppressed by an 
$\alpha_{\rm QED}$ factor, but this is compensated by the
fact that they are more abundantly produced in air
showers than charm or bottom hadrons. An important
consequence of these unflavored meson 
contributions is that the atmospheric
neutrino to muon ratio is smaller
than the one obtained by other authors. 

We have also determined the flux of very energetic 
dimuons. These muon pairs would
be a background to long-lived CHAMPs
produced in the air or the ground and reaching a 
neutrino telescope from
horizontal directions. If a minimum angle between
the two muons (so that they can be resolved) is 
imposed, we have obtained
that the dominant sources are heavy hadron decays
and Drell-Yan processes.

We think that neutrino telescopes may provide 
opportunities that go beyond the study of neutrino 
interactions at ultrahigh energies. The study of
the muon flux from near-horizontal directions would 
provide 
important information about hadronic collisions
at energies and $q^2$ (forward direction) not
accessible in colliders, and it could even reveal 
signals of non-standard physics.

\subsection*{Acknowledgements}

The computer simulations were performed at the UGRGRID of the 
`Secci\'on de Supercomputaci\'on' of the University of Granada.
We are grateful to Paolo Lipari for discussions.
This work has been supported by MEC of Spain (FPA2006-05294
and FPA2008-03630-E/INFN), by the Italian Ministero 
dell'Universit\`a e della Ricerca Scientifica 
(COFIN-PRIN 2006 and CICYT-INFN 10485/2008)
and by Junta de Andaluc{\'\i}a (FQM-101, FQM-437 and FQM-03048).


\begin{thebibliography}{10}

\bibitem{Amsler:2008zzb}
  C.~Amsler {\it et al.}  [Particle Data Group],
  %``Review of particle physics,''
  Phys.\ Lett.\  B {\bf 667} (2008) 1.
  %%CITATION = PHLTA,B667,1;%%

\bibitem{Greisen:1960wc}
K.~Greisen,
%``Cosmic Ray Showers,''
Ann.\ Rev.\ Nucl.\ Part.\ Sci.\  {\bf 10} (1960) 63.
%%CITATION = ARNUA,10,63;%%

\bibitem{Lipari:1993hd}
P.~Lipari,
%``Lepton spectra in the earth's atmosphere,''
Astropart.\ Phys.\  {\bf 1} (1993) 195.
%%CITATION = APHYE,1,195;%%

\bibitem{aires} S.J. Sciutto, in Procs. of the 27th ICRC, Hamburg, 2001
(Copernicus Gesellschafr, Hamburg, 001), Vol. 1, p. 237; see also
http://www.fisica.unlp.edu.ar/auger/aires.

\bibitem{Heck:1998vt}
D.~Heck, G.~Schatz, T.~Thouw, J.~Knapp and J.~N.~Capdevielle,
{\it CORSIKA: A Monte Carlo code to simulate extensive air showers},
Report FZKA-6019 (1998), Forschungszentrum Karlsruhe; 
http://www-ik.fzk.de/corsika

\bibitem{Costa:2000jw}
  C.~G.~S.~Costa,
  %``The prompt lepton cookbook,''
  Astropart.\ Phys.\  {\bf 16} (2001) 193.
  %[arXiv:hep-ph/0010306].
  %%CITATION = APHYE,16,193;%%

\bibitem{Volkova:1987gh}
  L.~V.~Volkova, W.~Fulgione, P.~Galeotti and O.~Saavedra,
  %``PROMPT MUON PRODUCTION IN COSMIC RAYS,''
  Nuovo Cim.\  C {\bf 10} (1987) 465.
  %%CITATION = NUCIA,10C,465;%%

\bibitem{Bugaev:1998bi}
  E.~V.~Bugaev, A.~Misaki, V.~A.~Naumov, T.~S.~Sinegovskaya, 
  S.~I.~Sinegovsky and N.~Takahashi,
  %``Atmospheric muon flux at sea level, underground and underwater,''
  Phys.\ Rev.\  D {\bf 58} (1998) 054001
  %[arXiv:hep-ph/9803488].
  %%CITATION = PHRVA,D58,054001;%%

\bibitem{Gondolo:1995fq}
  M.~Thunman, G.~Ingelman and P.~Gondolo,
  %``Charm production and high energy atmospheric muon and neutrino fluxes,''
  Astropart.\ Phys.\  {\bf 5} (1996) 309.
  %[arXiv:hep-ph/9505417].
  %%CITATION = APHYE,5,309;%%

\bibitem{Zas:1992ci}
  E.~Zas, F.~Halzen and R.~A.~V\'azquez,
  %``High-energy neutrino astronomy: Horizontal air shower arrays versus
  %underground detectors,''
  Astropart.\ Phys.\  {\bf 1} (1993) 297.
  %%CITATION = APHYE,1,297;%%

\bibitem{Pasquali:1998ji}
  L.~Pasquali, M.~H.~Reno and I.~Sarcevic,
  %``Lepton fluxes from atmospheric charm,''
  Phys.\ Rev.\  D {\bf 59} (1999) 034020.
  %[arXiv:hep-ph/9806428].
  %%CITATION = PHRVA,D59,034020;%%

\bibitem{Gelmini:1999xq}
  G.~Gelmini, P.~Gondolo and G.~Varieschi,
  %``Prompt atmospheric neutrinos and muons: Dependence on the gluon
  %distribution function,''
  Phys.\ Rev.\  D {\bf 61} (2000) 056011.
  %[arXiv:hep-ph/9905377].
  %%CITATION = PHRVA,D61,056011;%%

\bibitem{Enberg:2008te}
  R.~Enberg, M.~H.~Reno and I.~Sarcevic,
  %``Prompt neutrino fluxes from atmospheric charm,''
  Phys.\ Rev.\  D {\bf 78} (2008) 043005.
  %[arXiv:0806.0418 [hep-ph]].
  %%CITATION = PHRVA,D78,043005;%%

\bibitem{Kochanov:2009rn}
  A.~A.~Kochanov, T.~S.~Sinegovskaya, S.~I.~Sinegovsky, A.~Misaki and N.~Takahashi,
  %``Atmospheric muon flux at the PeV scale,''
  arXiv:0906.3791 [astro-ph.HE].
  %%CITATION = ARXIV:0906.3791;%%

\bibitem{icecube}
J.~Ahrens {\it et al.}  [IceCube Collaboration],
%``Sensitivity of the IceCube detector to astrophysical sources of high
%energy muon neutrinos,''
Astropart.\ Phys.\  {\bf 20} (2004) 507.
%[arXiv:astro-ph/0305196].
%%CITATION = ASTRO-PH 0305196;%%

\bibitem{Berghaus:2009jb}
  P.~Berghaus, for the IceCube Collaboration,
  ``Muons in IceCube,''
  arXiv:0902.0021 [astro-ph.HE].
  %%CITATION = ARXIV:0902.0021;%%

\bibitem{Ando:2007ds}
  S.~Ando, J.~F.~Beacom, S.~Profumo and D.~Rainwater,
  %``Probing new physics with long-lived charged particles produced by
  %atmospheric and astrophysical neutrinos,''
  JCAP {\bf 0804} (2008) 029.
  %[arXiv:0711.2908 [hep-ph]].
  %%CITATION = JCAPA,0804,029;%%

\bibitem{Ahlers:2007js}
  M.~Ahlers, J.~I.~Illana, M.~Masip and D.~Meloni,
  %``Long-lived Staus from Cosmic Rays,''
  JCAP {\bf 0708} (2007) 008.
  %[arXiv:0705.3782 [hep-ph]].
  %%CITATION = JCAPA,0708,008;%%

\bibitem{Illana:2006xg}
  J.~I.~Illana, M.~Masip and D.~Meloni,
  %``New physics from ultrahigh energy cosmic rays,''
  Phys.\ Rev.\  D {\bf 75} (2007) 055002.
  %[arXiv:hep-ph/0611036].
  %%CITATION = PHRVA,D75,055002;%%

\bibitem{pythia}
  T.~Sj\"ostrand, S.~Mrenna and P.~Skands,
  ``PYTHIA 6.4 Physics and Manual,''
  JHEP {\bf 0605} (2006) 026.
  %[arXiv:hep-ph/0603175].

\bibitem{Martin:2003us}
  A.~D.~Martin, M.~G.~Ryskin and A.~M.~Stasto,
  %``Prompt neutrinos from atmospheric $c \bar{c}$ and $b \bar{b}$ production
  %and the gluon at very small x,''
  Acta Phys.\ Polon.\  B {\bf 34} (2003) 3273.
  %[arXiv:hep-ph/0302140].
  %%CITATION = APPOA,B34,3273;%%

\bibitem{Ryazhskaya:2005rp}
  O.~G.~Ryazhskaya, L.~V.~Volkova and G.~T.~Zatsepin,
  %``Possible charm production and direct muon contribution to EAS at very high
  %energies,''
  Int.\ J.\ Mod.\ Phys.\  A {\bf 20} (2005) 6971.
  %%CITATION = IMPAE,A20,6971;%%

\bibitem{Donnachie:1992ny}
  A.~Donnachie and P.~V.~Landshoff,
  %``Total cross-sections,''
  Phys.\ Lett.\  B {\bf 296} (1992) 227.
  %[arXiv:hep-ph/9209205].
  %%CITATION = PHLTA,B296,227;%%

\bibitem{Schuler:1993wr}
  G.~A.~Schuler and T.~Sj\"ostrand,
  %``Hadronic diffractive cross-sections and the rise of the total
  %cross-section,''
  Phys.\ Rev.\  D {\bf 49} (1994) 2257.
  %%CITATION = PHRVA,D49,2257;%%

\bibitem{Andersson:1983ia}
  B.~Andersson, G.~Gustafson, G.~Ingelman and T.~Sj\"ostrand,
  %``Parton Fragmentation And String Dynamics,''
  Phys.\ Rept.\  {\bf 97} (1983) 31.
  %%CITATION = PRPLC,97,31;%%

\bibitem{Pasquali:1998xf}
  L.~Pasquali and M.~H.~Reno,
  %``Tau neutrino fluxes from atmospheric charm,''
  Phys.\ Rev.\  D {\bf 59} (1999) 093003.
  %[arXiv:hep-ph/9811268].
  %%CITATION = PHRVA,D59,093003;%%

\bibitem{Costa:2001fb}
  C.~G.~S.~Costa, F.~Halzen and C.~Salles,
  %``The prompt TeV-PeV atmospheric neutrino window,''
  Phys.\ Rev.\  D {\bf 66} (2002) 113002.
  %[arXiv:hep-ph/0104039].
  %%CITATION = PHRVA,D66,113002;%%

\bibitem{Lee:2004zm}
  F.~F.~Lee and G.~L.~Lin,
  %``A Semi-analytic calculation on the atmospheric 
  %tau neutrino flux in the GeV
  %to TeV range,''
  Astropart.\ Phys.\  {\bf 25} (2006) 64.
  %[arXiv:hep-ph/0412383].
  %%CITATION = APHYE,25,64;%%

\bibitem{Albuquerque:2006am}
  I.~F.~M.~Albuquerque, G.~Burdman and Z.~Chacko,
  %``Direct Detection of Supersymmetric Particles in Neutrino Telescopes,''
  Phys.\ Rev.\  D {\bf 75} (2007) 035006.
  %[arXiv:hep-ph/0605120].
  %%CITATION = PHRVA,D75,035006;%%

\bibitem{GonzalezGarcia:2006ay}
  M.~C.~Gonz\'alez-Garc{\'\i}a, M.~Maltoni and J.~Rojo,
  %``Determination of the atmospheric neutrino fluxes from atmospheric  neutrino
  %data,''
  JHEP {\bf 0610} (2006) 075.
  %[arXiv:hep-ph/0607324].
  %%CITATION = JHEPA,0610,075;%%

\end{thebibliography}
\end{document}